%% file: hsc_sn.tex
\documentclass[article]{pasj01}
 \jyear{2019}
\usepackage{url}

\begin{document}

\newcommand{\comment}[1]{\textcolor{red}{\textbf{#1}}}
\newcommand{\ncont}{9219}

\title{Searches for Population III Pair-Instability Supernovae: Impact of Gravitational Lensing Magnification}

\author{Kenneth C. \textsc{Wong}\altaffilmark{1,2}}
\email{ken.wong@ipmu.jp}

\author{Takashi J. \textsc{Moriya}\altaffilmark{2}}

\author{Masamune \textsc{Oguri}\altaffilmark{1,3,4}}

\author{Stefan \textsc{Hilbert}\altaffilmark{5,6}}

\author{Yusei \textsc{Koyama}\altaffilmark{7}}

\author{Ken'ichi \textsc{Nomoto}\altaffilmark{1}}

\altaffiltext{1}{Kavli IPMU (WPI), UTIAS, The University of Tokyo, Kashiwa, Chiba 277-8583, Japan}
\altaffiltext{2}{National Astronomical Observatory of Japan, 2-21-1 Osawa, Mitaka, Tokyo 181-8588, Japan}
\altaffiltext{3}{Department of Physics, The University of Tokyo, 7-3-1 Hongo, Bunkyo-ku, Tokyo 113-0033, Japan}
\altaffiltext{4}{Research Center for the Early Universe, University of Tokyo, Tokyo 113-0033, Japan}
\altaffiltext{5}{Exzellenzcluster Universe, Boltzmannstr. 2, 85748 Garching, Germany}
\altaffiltext{6}{Ludwig-Maximilians-Universit{\"a}t, Universit{\"a}ts-Sternwarte, Scheinerstr. 1, 81679 M{\"u}nchen, Germany}
\altaffiltext{7}{Subaru Telescope, National Astronomical Observatory of Japan, National Institutes of Natural Sciences, 650 North A'ohoku Place, Hilo, HI 96720, USA}


\maketitle

\begin{abstract}
Superluminous supernovae have been proposed to arise from Population III progenitors that explode as pair-instability supernovae.  Pop III stars are the first generation of stars in the Universe, and are thought to form as late as $z \sim 6$.  Future near-infrared imaging facilities such as ULTIMATE-Subaru can potentially detect and identify these PISNe with a dedicated survey.  Gravitational lensing by intervening structure in the Universe can aid in the detection of these rare objects by magnifying the high-$z$ source population into detectability.  We perform a mock survey with ULTIMATE-Subaru, taking into account lensing by line-of-sight structure to evaluate its impact on the predicted detection rate.  We compare a LOS mass reconstruction using observational data from the Hyper Suprime Cam survey to results from cosmological simulations to test their consistency in calculating the magnification distribution in the Universe to high-$z$, but find that the data-based method is still limited by an inability to accurately characterize structure beyond $z \sim1.2$.  We also evaluate a survey strategy of targeting massive galaxy clusters to take advantage of their large areas of high magnification.  We find that targeting clusters can result in a gain of a factor of $\sim$two in the predicted number of detected PISNe at $z > 5$, and even higher gains with increasing redshift, given our assumed survey parameters.  For the highest-redshift sources at $z \sim 7-9$, blank field surveys will not detect any sources, and lensing magnification by massive clusters will be necessary to observe this population.
\end{abstract}

\section{Introduction} \label{sec:intro}
Recent optical transient surveys have uncovered a population of supernovae (SNe) that are extremely luminous.  These superluminous supernovae (SLSNe; e.g., \cite{quimby+2011,galyam+2012}) are often more than 10 times brighter than ordinary SNe.  Various theoretical models have been proposed to explain the extreme luminosities of these objects (see \cite{moriya+2018} for a review).  One such proposed explanation is that the brightness of SLSNe is due to the production of large quantities of radioactive $^{56}$Ni in a ``pair instability" supernova (PISN; e.g., \cite{smith+2007,galyam+2009b}), the explosion of a very massive star with a helium core between $70~\mathrm{M_{\odot}} \lesssim \mathrm{M} \lesssim 140~\mathrm{M_{\odot}}$ \citep{heger+2002}.  Stars with such a high core mass are thought to present at metallicities lower than one third of solar in order to keep their mass long enough to explode as PISNe \citep{langer+2007}.

The most promising candidate progenitors of PISNe are Population III (Pop III) stars, which are the first generation of stars in the Universe.  Due to their low metallicity, they do not suffer from wind mass loss and can grow cores massive enough to explode as PISNe.  Pop III stars are thought to form as late as $z \sim 6$ (e.g., \cite{desouza+2014}).  Previous studies have attempted to estimate the detectability of high-redshift PISNe resulting from the explosions of these stars in upcoming imaging surveys such as {\it Euclid} \citep{laureijs+2011} and the Wide Field Infrared Survey Telescope (WFIRST).

Gravitational lensing by intervening structure along the line of sight (LOS) to SNe can potentially aid in the detectability of such events due to lensing magnification (e.g., \cite{quimby+2014}).  Past studies (e.g., \cite{tanaka+2012,tanaka+2013,whalen+2013}) have generally ignored this effect.  The distribution of magnifications in the Universe has typically been calculated from cosmological simulations (e.g., \cite{hilbert+2007,hilbert+2008,lima+2010,takahashi+2011,castro+2018}).  \citet{kronborg+2010} calculated magnifications along specific lines of sight to $z \sim 1$ type Ia SNe, normalized by the median to random lines of sight in the Supernova Legacy Survey (SNLS), but reconstructing the magnification distribution using observational data has generally not been widely attempted (although see \cite{sakakibara+2019}), particularly to high redshifts.  With the current state-of-the-art wide-area imaging surveys, these type of reconstructions may be more feasible than they were with past datasets.

In this paper, along with a companion paper (Moriya et al. 2019; hereafter M19), we estimate how many $z \gtrsim 5$ PISNe could be detected in future surveys by performing mock observations with two upcoming facilities: the Ultra-wide Laser Tomographic Imager and MOS with AO for Transcendent Exploration by Subaru Telescope (ULTIMATE-Subaru), and WFIRST.  We account for the influence of lensing by intervening structure, as well as investigate the benefits of a survey strategy that targets massive clusters to take advantage of their gravitational lensing properties.  In M19, we focus on survey strategy and observational methods to maximize the number of high-$z$ PISNe candidates found with these upcoming facilities.  In this paper, we describe the details of our lensing calculations, including a comparison between magnification distributions derived from observational data and simulations, which will inform future studies using lensing magnification to study faint, rare phenomena.  We focus on one particular survey strategy with ULTIMATE-Subaru in this paper, as its field of view is better suited to a direct comparison of a blank field survey and a cluster survey, and refer the reader to M19 for expanded results for WFIRST and different survey strategies.

This paper is organized as follows.  We describe the optical imaging data and the cosmological simulations used in this study in Section~\ref{sec:data}.  We describe a mock survey with ULTIMATE-Subaru to estimate the detectability of high-redshift PISNe accounting for lensing effects in Section~\ref{sec:sn}.  In Section~\ref{sec:massmod}, we describe our method for determining the magnification distribution in the universe from the observational data in order to compare with the distribution from cosmological simulations.  We also explore a strategy of targeting massive clusters to exploit their large areas of high magnification.  We present our main results in Section~\ref{sec:results} and summarize our conclusions in Section~\ref{sec:conclusion}.  Throughout this paper, we assume $\Omega_{m} = 0.3$, $\Omega_{\Lambda} = 0.7$, and $h = 0.7$.  All magnitudes given are on the AB system.

\section{Data} \label{sec:data}

\subsection{Hyper Suprime-Cam Imaging Data} \label{subsec:hsc}
The Hyper Suprime-Cam Subaru Strategic Program (HSC SSP; \cite{aihara+2018a}) is an ongoing imaging survey with the Hyper Suprime-Cam (HSC; \cite{miyazaki+2012,miyazaki+2018,furusawa+2018,kawanomoto+2018,komiyama+2018}) on the Subaru Telescope.  The depth and image quality of the HSC SSP makes it an idea dataset with which to attempt to reconstruct the magnification distribution along the LOS to high-$z$ sources.  The Wide component of the HSC SSP will observe a $\sim1400$ deg$^{2}$ area in the $grizy$ bands to a depth of $i = 26.2$.  The data used in this study come from Data Release 2 (hereafter DR2) of the HSC SSP, which includes data taken through the S17A semester covering 776 deg$^{2}$ in all bands, including 289 deg$^{2}$ to the full depth.  The median $i$-band seeing of the data is $\sim0\farcs6$.  The data are reduced with {\tt hscPipe} version 5.4.0 \citep{bosch+2018}, which is based on the Large Synoptic Survey Telescope (LSST) pipeline \citep{axelrod+2010,juric+2015}.

The galaxy magnitudes used in this analysis are cModel magnitudes, which are measured by fitting galaxy models convolved with the point spread function (PSF) to the light profile of the object \citep{abazajian+2004}.  Photometric redshifts and stellar masses are determined using the {\sc mizuki} algorithm \citep{tanaka2015}.  A description of its application to the HSC SSP data is presented in \citet{tanaka+2018}.  The robustness of the photometric redshifts is a function of galaxy redshift and brightness, and are quantified in terms of $\Delta z / (1+z_{\mathrm{ref}})$, where $\Delta z \equiv |z - z_{\mathrm{ref}}|$ and $z_{\mathrm{ref}}$ is a reference redshift.  We only use galaxies that are brighter than $i = 24$ and that have a photometric redshift of $0.2 \leq z \leq 1.2$, as the photo-z accuracy drops off substantially beyond these limits.  We require that all galaxies are observed in all five HSC bands, even if not to the full depth, as this affects the photometric redshift accuracy.

\subsection{CAMIRA Cluster Catalog} \label{subsec:camira}
We account for group and cluster halos in the HSC SSP data using the catalog of \citet{oguri+2018}, which uses the CAMIRA algorithm \citep{oguri2014} to select halos based on a multiband identification of red sequence galaxies.  CAMIRA also estimates a richness (defined as the number of red member galaxies with stellar masses $\mathrm{M_{*} \gtrsim 10^{10.2}~M_{\odot}}$ and projected within $R \lesssim 1~h^{-1}$ Mpc) and a photometric redshift for each halo. We use an expanded version of the DR2 catalog that includes 13876 groups and clusters with richness $> 10$ and photometric redshifts ranging from $0.1 < z < 1.4$.  In cases where the brightest cluster galaxy (BCG) has a measured spectroscopic redshift, we take that value to be the cluster redshift.

\subsection{Cosmological Simulations} \label{subsec:sims}
We use the results of ray tracing calculations through the Millennium simulation \citep{springel+2005} to calculate the magnification distribution in the Universe and to compare to the results based on the observational HSC data. The details of these calculations are described in \citet{hilbert+2007}, \citet{hilbert+2008}, and \citet{hilbert+2009}.

\begin{figure}
\begin{center}
\includegraphics[width=8cm]{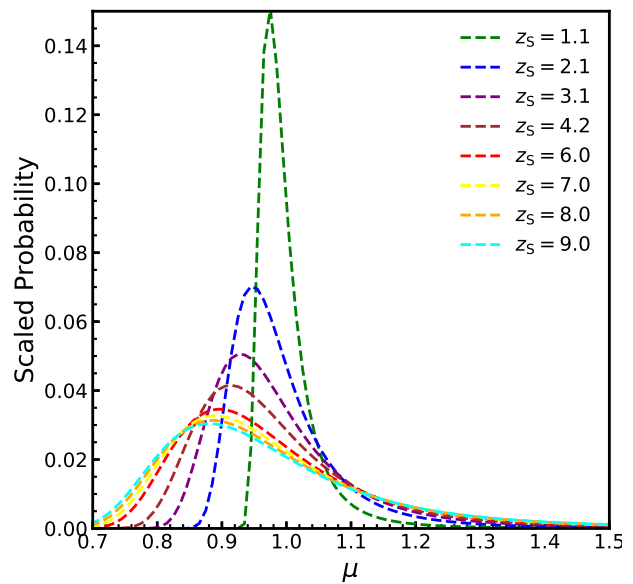}
\end{center}
\caption{
Source plane magnification distribution from the simulations of \citet{hilbert+2008} for a variety of source redshifts ranging from $z_{\mathrm{S}} \sim 1$ to $z_{\mathrm{S}} \sim 9$.  The simulations include both dark matter and baryons.  In the case of multiple imaging, the magnification of the brightest image is used.
} \label{fig:sims_maghist}
\end{figure}

The dark-matter particle distribution in the Millennium simulation is used to populate 32 lightcones, each with a $4\times 4\,~\mathrm{deg}^2$ field of view. The lightcones are divided into redshift slices whose matter content is projected onto lens planes. In addition to the simulation particles, assumed to represent the dark matter and gas, the stellar mass in galaxies as predicted by semi-analytic models of galaxy formation \citep{delucia+2007} is represented by analytic mass profiles. For each light cone, $4096^2$ light rays are traced back from the observer through the 30 (for redshifts $z \sim 2$) to 50  (for $z \sim 10$) lens planes, and the resulting magnification values are used to estimate the magnification distribution as a function of source redshift.  Figure~\ref{fig:sims_maghist} shows the magnification distributions determined from these simulations for source redshifts ranging from $z_{\mathrm{S}} \sim 1$ to $z_{\mathrm{S}} \sim 9$.

\section{Detectability of Supernovae} \label{sec:sn}

\subsection{Mock Survey with ULTIMATE-Subaru} \label{subsec:survey}
We set up a mock transient survey with ULTIMATE-Subaru to evaluate the detectability of high-z PISNe, accounting for lensing magnification.  ULTIMATE-Subaru is an upcoming wide-field adaptive optics system with a NIR instrument currently planned to have a field of view of $14\arcmin \times 14\arcmin$.  The details of the instrument have not been finalized, but it is expected that the reddest filter available will be the $K$-band.  For our calculations, we assume the $K$-band filter and throughput of the Multi-Object Infrared Camera and Spectrograph (MOIRCS; \cite{ichikawa+2006,suzuki+2008}) on Subaru, as it is likely to be close to similar ULTIMATE-Subaru observations.  We assume standard seeing conditions of 0.2\arcsec~within a 0.3\arcsec~aperture.

Our mock 5-year survey assumes an observing interval of $\mathrm{t_{int}} = 180$ days, which works well to find Pop III PISNe at $z \gtrsim 6$.  We set a minimum number of detections $\mathrm{N_{d}} = 2$ for a transient to be considered a discovery.  We assume a total survey area of 1 deg$^{2}$, which would require $\sim 18$ pointings of ULTIMATE-Subaru centered on clusters with average magnification properties similar to the clusters we use in our calculations (Section~\ref{subsec:cluster}).  We perform calculations for survey depths of $K = 26.0$ and $K = 26.5$ mag (adopting a signal-to-noise ratio of 5), and for source redshifts ranging from $z > 5$ to $z > 9$.  With these assumed observing parameters, our mock survey would require 480 hours and 860 hours of exposure time, respectively, over the five year period to obtain depths of $K = 26.0$ and $K = 26.5$.  In M19, we also discuss alternative survey strategies with ULTIMATE-Subaru and WFIRST that have a larger survey area, shorter $\mathrm{t_{int}}$, or higher $\mathrm{N_{d}}$, so we refer the reader to that work for expanded results.

\subsection{PISN Properties} \label{subsec:pisne}
The details of our assumed PISN properties are presented in M19, but we provide a brief summary here.  We adopt PISN light curves predicted by \citet{kasen+2011}, which are numerically obtained from the PISN progenitors of \citet{heger+2002}.  The peak luminosity of the R250 model exceeds $K = 26.0$ mag at $z =6$, while the R225 model is brighter than $K = 26.5$ mag at $z = 6$.  Thus, the R225 explosions at $z \gtrsim 6$ can only be observed when we conduct the transient surveys with a limiting magnitude of 26.5 mag in the $K$-band.  The R225 and R250 models have red supergiant (RSG) progenitors.  There are other models that have hydrogen-free PISN progenitors, but it is not obvious from which zero-age main sequence (ZAMS) masses these bare helium core progenitors originate. Therefore, we perform mock observations only considering Pop III RSG PISNe progenitors.

We adopt a PISN rate estimated by \citet{desouza+2014}, which is based on the cosmological simulation of \citet{johnson+2013}.  We take the Pop III PISNe rates from their SFR10 model, which has a relatively high SFR prediction compared to others.  Cosmological simulations of the first stars indicate that the initial mass function (IMF) of Pop III stars is close to flat \citep{hirano+2015}, so we assume a flat IMF for our calculations, although M19 also show results for an alternative IMF.

\section{Lensing Calculations and Mass Models} \label{sec:massmod}

\subsection{Magnification by LOS Structure} \label{subsec:lss}
To estimate the distribution of magnifications in the Universe, we generate mass distributions in random fields from the HSC SSP.  The fields are chosen by randomly selecting $12\arcmin \times12\arcmin$ patches from the survey catalog, then selecting random coordinates within that patch.  The random fields are circular apertures with a radius of 120\arcsec.  There are gaps within the survey region due to chip gaps, bright star masks, etc., which can bias our results.  To mitigate this effect, we use the HSC SSP random point catalog \citep{coupon+2018}, which contains randomly sampled points in the survey region that are flagged in the same way as the objects.  The random points are drawn with a density of 100 points per arcmin$^{2}$.  We remove fields that contain $\leq 95\%$ of the number of points expected from a completely unmasked field to ensure that masking is not a significant issue.  Our final sample comprises \ncont~fields from the HSC SSP.

Within each field, we assign mass to galaxies.  The photometric redshifts from {\sc mizuki} are accurate between $0.2 \lesssim z \lesssim 1.2$ \citep{tanaka+2018}, so we ignore galaxies outside of this redshift range.  At $z \leq 0.2$, there is relatively little volume, so we expect this to be a small effect.  Excluding galaxies at $z \geq 1.2$ may result in neglecting a non-trivial amount of matter to high source redshifts, so we compare our calculation to results from cosmological simulations (Section~\ref{subsec:sims}) to determine whether this is valid.  We assume the galaxies are located at their best-fit photometric redshift as determined by {\sc mizuki}.  We neglect photometric redshift uncertainty, as it is subdominant to the uncertainty in the stellar masses.  We remove galaxies with best-fit stellar masses $\mathrm{M_{*} \leq 10^{8}~M_{\odot}}$, as these galaxies are generally faint and difficult to characterize accurately.

We assign mass to galaxies by taking their best-fit stellar mass from the HSC catalog and calculating their halo mass using the stellar mass-halo mass relation (SHMR) of \citet{leauthaud+2012}.  Uncertainty in the stellar mass measurements are accounted for by drawing a halo mass from a Gaussian distribution with a width equal to the geometric mean of the 16\% and 84\% quantiles.  Scatter in the SHMR is accounted for in a similar way.  Due to the steepness of the SHMR at the high-stellar mass end, we set an upper limit of $M_{\mathrm{h}} = 10^{13.5}~\mathrm{M_{\odot}}$ to prevent unphysically large halos from being assigned to individual galaxies.  The galaxies are assumed to be singular isothermal spheres (SIS) with a truncation radius of $r_{200\mathrm{m}}$, at which the mean density inside a sphere of that radius centered on the galaxy is 200 times the mean matter density of the universe at the galaxy's redshift.

Group and cluster halos identified in the survey are included as spherical NFW \citep{navarro+1997} halos.  Since their influence can extend over greater projected distances, we include these larger halos if they are within 300\arcsec~of a field center.  We do not account for uncertainty in the group and cluster photometric redshifts from the CAMIRA algorithm, as these effects on the global magnification distribution should wash out over a large number of random fields.  The halo masses are estimated from their richness \citep{oguri+2018}, and the halo concentration is calculated using the results of simulations by \citet{zhao+2009}.

We calculate the magnification at the center of each field for a range of source redshifts using an updated version of the {\sc gravlens} software \citep{keeton2001}, which assumes that all of the mass in the structures we include explicitly are added to a line of sight containing a smooth matter distribution equal to the mean density in the Universe.  The mean magnification in the universe must be unity in order to conserve flux, so we shift these uncorrected magnifications, $\mu^{\prime}$, by an amount $\mu_\mathrm{c}$ such that $\langle \mu^{\prime}-\mu_\mathrm{c} \rangle = 1$.  We then calculate the distribution of true magnifications, $\mu = \mu^{\prime}-\mu_\mathrm{c}$, weighted by $(\mu^{\prime}-\mu_\mathrm{c})^{-1}$ to get a uniform distribution of sources in the source plane.

To increase the speed of the calculations, we use the tidal approximation of \citet{mccully+2014} for most of the galaxies in the field.  All group and cluster halos are treated exactly, as are galaxies that are either projected within 60\arcsec~of the field center or are assigned a halo mass of $M_{\mathrm{h}} \geq 10^{12}~\mathrm{M_{\odot}}$.  We verify for a subset of fields that this approximation yields magnification results that are nearly identical with the results when all structures are explicitly included in the mass model.

\subsection{Magnification by Massive Galaxy Clusters} \label{subsec:cluster}
In addition to evaluating the effect of lensing on a survey of random lines of sight, we also investigate a strategy of targeting massive clusters in order to take advantage of lensing magnification, which can potentially magnify high-$z$ SNe into detectability.  An ideal survey to maximize the cross-section of intermediate-to-high magnification regions would target the most massive known galaxy clusters, or lines of sight with multiple massive structures \citep{wong+2012,french+2014}.

To estimate the number of detections of high-$z$ SNe with such a survey, we use existing models of seven known massive clusters to determine their magnification distributions for point sources at high redshift, then extrapolate these results to our mock survey (Section~\ref{sec:sn}).  We use the model of the massive cluster J0850+3604 \citep{wong+2017}, as well as models of the six {\it HST} Frontier Fields (HFF; \cite{lotz+2017}) clusters.  This is somewhat optimistic, as these are already among the best lensing clusters that are known and that have been characterized.  However, there may be other clusters with similarly good magnification properties that are unexplored (e.g., \cite{wong+2013}), and current and future wide-area imaging surveys such as HSC, LSST, and Euclid could potentially find and identify many more.

The mass model we use for J0850+3604 is the best-fit model from \citet{wong+2017}, while the HFF cluster models are taken from \citet{kawamata+2016} and \citet{kawamata+2018} using the {\tt glafic} code \citep{oguri2010}.  Both methods use parameterized models account for both the cluster dark matter distribution and individual galaxies.  We use single models without accounting for uncertainties, as the impact on the integrated magnification distribution for each cluster is small compared to the sample variance.  We note that the J0850+3604 model accounts for line-of-sight structure while the HFF models do not, but we re-run the magnification calculation with only the mass at the redshift of the massive J0850+3604 cluster included and find that it has a negligible effect.

We assume a $14\arcmin \times 14\arcmin$ field of view, which is roughly the expected field of view for ULTIMATE-Subaru.  We calculate the source plane area as a function of magnification for each of the seven clusters and take the average, extrapolated to the full area of our mock survey, to calculate the expected number of detections.  For regions of the source plane that are multiply-imaged, we take the magnification to be that of the brightest image, as that is the relevant quantity for detectability.

\section{Results} \label{sec:results}

\subsection{Comparison of LOS Magnification from Data vs. Simulations} \label{subsec:los_mag}
\begin{figure}
\begin{center}
\includegraphics[width=8cm]{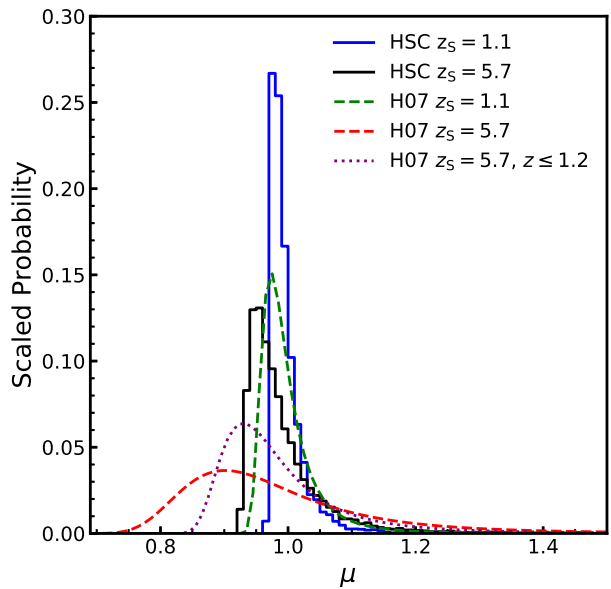}
\end{center}
\caption{
Source plane magnification distribution calculated from HSC SSP data (solid lines) compared with magnification distribution from the simulations of \citet{hilbert+2008} (dashed lines).  In the case of multiple imaging, the magnification of the brightest image is used.  The HSC magnification distribution at $z_{\mathrm{S}} = 1.1$ is more narrow than the distribution from the simulations.  At $z_{\mathrm{S}} = 5.7$, the distribution from simulations is much broader and peaks at a lower magnification than that from observational data, suggesting that the matter distribution beyond $z = 1.2$ is an important contributor to the magnification distribution.  We compare to the simulation results, only including matter up to $z = 1.2$ (dotted line), which moves the distributions closer.
} \label{fig:magdist}
\end{figure}

Figure~\ref{fig:magdist} shows the magnification distribution derived from the HSC SSP data compared with that derived from the simulations of \citet{hilbert+2008}.  We show the distributions for source redshifts of $z_{\mathrm{S}} = 1.1$ and $z_{\mathrm{S}} = 5.7$ to match the \citet{hilbert+2008} study, and to highlight the limitations of the data.

The first result to note here is that the magnification distribution for $z_{\mathrm{S}} = 1.1$ based on HSC data is more narrow and peaks slightly higher than the distribution from simulations.  Since this is within the redshift range where we are still accounting for galaxies and clusters, and the volume within $z \leq 0.2$ is a small fraction of the total volume, this could arise from various effects in our characterization of the LOS mass distribution.  One possibility is that there may be a systematic bias in the stellar mass determinations or in the stellar-to-halo mass scaling.  In particular, we find much closer agreement between the HSC calculation and the \citet{hilbert+2008} simulation results when arbitrarily scaling up the galaxy stellar masses, although the DR2 stellar masses from {\sc mizuki} have been tested and are consistent with a training sample derived from a more extensive multi-band dataset (M. Tanaka, private communication).  There may also be non-negligible effects from structures smaller than $\mathrm{M_{*} \leq 10^{8}~M_{\odot}}$.  Although the HSC SSP represents the current best wide-field survey to characterize the mass distribution along individual lines of sight to significant depth, we caution that this method is still limited by systematic effects that need to be further explored, even for calculations at lower source redshifts.

Another issue is that the distributions for $z_{\mathrm{S}} = 5.7$ significantly disagree.  The distribution from simulations is much broader in comparison to the one derived from data, much more so than the $z_{\mathrm{S}} = 1.1$ results.  This suggests that structure beyond $z = 1.2$, which is excluded from the HSC results, is an important contributor to the magnification distribution, which is not necessarily obvious because bound structures at higher redshift are generally less massive than those at lower redshift, and the lensing efficiency of matter beyond $z \sim 1$ is continuously decreasing (regardless of source redshift) due to the increasing critical surface density for lensing.  If we compare the results to those of simulations only including matter up to $z = 1.2$, the distributions become closer, but the observational results are still more narrow and peak at higher values, similar to the $z = 1.1$ results.  This indicates that using observational data to constrain the magnification distribution to high redshifts, in addition to having potential systematic effects mentioned above, will likely be limited until we have the depth to perform an accurate reconstruction of the mass distribution out to the source redshift, as bound structures beyond $z \sim 1$ are non-negligible and cannot be approximated by a uniform density LOS.  Throughout the rest of this paper, we present results for lensing by LOS structure using the simulation results from \citet{hilbert+2008}.

\subsection{Impact of Lensing by LOS Structure on Detectability of High-z PISNe} \label{subsec:sne_lensing}
In Table~\ref{tab:sne_counts}, we show the predicted numbers of PISNe discovered in our mock survey across a range of redshifts ranging from $z > 5$ to $z > 9$.  We first examine the case of a random field survey, both with and without lensing magnification by intervening structure.  Comparing these two cases, we see that lensing does not make a meaningful difference in the expected number of detections, as the large majority of the source plane area has a magnification close the $\mu \sim 1$.

\renewcommand*\arraystretch{1.2}
\input{tab_sne_counts.tex}
\renewcommand*\arraystretch{1.0}

\subsection{Boost from Lensing by Massive Clusters} \label{subsec:sne_cluster}
We calculate the number of expected detections of PISNe for observations targeting each of the seven cluster models.  Table~\ref{tab:sne_counts} shows these numbers for each cluster, extrapolating the counts to the full 1 deg$^{2}$ of our mock survey.  We also report the average over the clusters (again extrapolated to 1 deg$^{2}$), which should be more representative of what could be expected from a real survey of 18 different clusters with similar properties.

These results suggest that a survey targeting massive clusters can lead to significant gains in the numbers of detected PISNe, particularly at high source redshifts, as has been previously suggested (e.g., \cite{tanaka+2013,whalen+2013}).  At $z > 5$, the gains can be a factor of $\sim$two or more, depending on the survey depth.  For higher redshift sources, although the total number of detection decreases, the fractional gain provided by lensing clusters is larger, particularly at $z > 7$, where a random field survey is not expected to detect any sources.

\section{Conclusions} \label{sec:conclusion}
We have investigated the impact of gravitational lensing on the detectability of high-redshift ($z \gtrsim 5$) pair-instability supernovae from future transient surveys, with a focus on the upcoming ULTIMATE-Subaru instrument.  Using data from the HSC SSP, we have attempted to reconstruct the magnification distribution in the Universe from observational data.  Despite the depth of the survey data, we are unable to accurately characterize the matter distribution beyond $z = 1.2$.  This, along with other sources of uncertainty, leads to disagreements with the magnification distributions predicted from cosmological simulations.

Using the simulation results from \citet{hilbert+2008}, we set up a mock transient survey with ULTIMATE-Subaru to evaluate the impact of lensing by large scale structure on high-$z$ PISNe.  We find that for a survey of random fields, lensing has a negligible impact on the predicted number of detections.  We also evaluate the benefits of a survey strategy that specifically targets massive galaxy clusters to take advantage of lensing magnification.  Using published mass models of seven massive clusters, we calculate the predicted number of detections in these fields, taking the average as a reasonable prediction for a future survey of the most massive known clusters. This strategy can increase the number of detections by a factor of $\sim$two, increasing to higher source redshifts at which blank field surveys would not detect any sources.

In a companion paper, M19, we apply these methods to different survey strategies and WFIRST observations, with suggestions regarding survey strategy and follow-up confirmation.  As new upcoming facilities such as ULTIMATE-Subaru and WFIRST come online, combining their observational power with the natural cosmic telescopes provided by gravitational lensing will allow us to observe the earliest phenomena from the first generation of stars.

\begin{ack}
We thank the referee, whose comments and suggestions were helpful in improving this paper.  We thank Masayuki Tanaka for providing insight into the stellar mass calculation from the {\sc mizuki} algorithm for HSC SSP data.  We thank Atsushi Nishizawa for helpful discussions and feedback.
The Hyper Suprime-Cam (HSC) collaboration includes the astronomical
communities of Japan and Taiwan, and Princeton University. The HSC instrumentation
and software were developed by the National Astronomical
Observatory of Japan (NAOJ), the Kavli Institute for the Physics and
Mathematics of the Universe (Kavli IPMU), the University of Tokyo, the
High Energy Accelerator Research Organization (KEK), the Academia
Sinica Institute for Astronomy and Astrophysics in Taiwan (ASIAA), and
Princeton University. Funding was contributed by the FIRST program
from Japanese Cabinet Office, the Ministry of Education, Culture, Sports,
Science and Technology (MEXT), the Japan Society for the Promotion of
Science (JSPS), Japan Science and Technology Agency (JST), the Toray
Science Foundation, NAOJ, Kavli IPMU, KEK, ASIAA, and Princeton
University.
Based in part on data collected at the Subaru Telescope and retrieved from the HSC data archive system, which is operated by the Subaru Telescope and Astronomy Data Center at National Astronomical Observatory of Japan.
This work was supported by World Premier International Research Center Initiative (WPI Initiative), MEXT, Japan.
K.C.W. is supported in part by an EACOA Fellowship awarded by the East Asia Core Observatories Association, which consists of the Academia Sinica Institute of Astronomy and Astrophysics, the National Astronomical Observatory of Japan, the National Astronomical Observatories of the Chinese Academy of Sciences, and the Korea Astronomy and Space Science Institute.
T. J. M. is supported by the Grants-in-Aid for Scientific Research of the Japan Society for the Promotion of Science (16H07413, 17H02864, 18K13585).
M.O. is supported in part by JSPS KAKENHI Grant Number JP15H05892, JP18K03693, and JP18H04572.
\end{ack}

\bibliographystyle{myaasjournal}
\bibliography{hsc_sn}

\end{document}

%% file: tab_sne_counts.tex
\begin{table*}
\caption{Predicted Numbers of RSG Pop~III PISN Discoveries for ULTIMATE-Subaru Mock Survey}\label{tab:sne_counts}
\begin{tabular}{l|ccccc}
source & $z > 5$ & $z > 6$ & $z > 7$ & $z > 8$ & $z > 9$ \\
\hline
&\multicolumn{5}{c}{$K=26.5$ mag limit} \\
random field (no lensing) & $7.89 \pm 0.39$ & $2.42 \pm 0.20$ & $0.06 \pm 0.03$ & $0.00 \pm 0.00$ & $0.00 \pm 0.00$
 \\
random field (lensing) & $8.13 \pm 0.37$ & $2.93 \pm 0.22$ & $0.70 \pm 0.10$ & $0.19 \pm 0.06$ & $0.08 \pm 0.04$
 \\
Abell~370 & $14.96 \pm 0.53$ & $8.85 \pm 0.40$ & $5.68 \pm 0.31$ & $2.60 \pm 0.22$ & $1.49 \pm 0.17$
 \\
Abell~1063 & $11.56 \pm 0.46$ & $5.82 \pm 0.32$ & $3.09 \pm 0.27$ & $1.49 \pm 0.18$ & $0.91 \pm 0.13$
 \\
Abell~2744 & $10.53 \pm 0.42$ & $4.93 \pm 0.30$ & $2.33 \pm 0.20$ & $1.06 \pm 0.13$ & $0.65 \pm 0.10$
 \\
J0416 & $10.21 \pm 0.40$ & $4.65 \pm 0.26$ & $2.10 \pm 0.18$ & $0.92 \pm 0.11$ & $0.54 \pm 0.10$
 \\
J0717 & $14.20 \pm 0.47$ & $8.29 \pm 0.35$ & $5.33 \pm 0.30$ & $2.99 \pm 0.24$ & $1.98 \pm 0.19$
 \\
J0850 & $12.32 \pm 0.47$ & $6.40 \pm 0.34$ & $3.47 \pm 0.26$ & $1.34 \pm 0.14$ & $0.67 \pm 0.10$
 \\
J1149 & $11.25 \pm 0.42$ & $5.58 \pm 0.27$ & $2.86 \pm 0.21$ & $1.31 \pm 0.15$ & $0.77 \pm 0.12$
 \\
cluster average & $12.15 \pm 1.68$ & $6.36 \pm 1.50$ & $3.55 \pm 1.31$ & $1.67 \pm 0.74$ & $1.00 \pm 0.49$
 \\
\hline
&\multicolumn{5}{c}{$K=26.0$ mag limit} \\
random field (no lensing) & $2.44 \pm 0.20$ & $0.52 \pm 0.08$ & $0.00 \pm 0.00$ & $0.00 \pm 0.00$ & $0.00 \pm 0.00$
 \\
random field (lensing) & $2.93 \pm 0.24$ & $0.61 \pm 0.09$ & $0.08 \pm 0.03$ & $0.04 \pm 0.02$ & $0.02 \pm 0.02$
 \\
Abell~370 & $7.17 \pm 0.36$ & $3.20 \pm 0.24$ & $1.65 \pm 0.17$ & $1.02 \pm 0.14$ & $0.71 \pm 0.13$
 \\
Abell~1063 & $5.10 \pm 0.30$ & $2.02 \pm 0.19$ & $1.00 \pm 0.14$ & $0.65 \pm 0.11$ & $0.45 \pm 0.09$
 \\
Abell~2744 & $4.54 \pm 0.29$ & $1.60 \pm 0.17$ & $0.72 \pm 0.11$ & $0.49 \pm 0.09$ & $0.34 \pm 0.08$
 \\
J0416 & $4.21 \pm 0.28$ & $1.39 \pm 0.17$ & $0.58 \pm 0.09$ & $0.37 \pm 0.07$ & $0.26 \pm 0.06$
 \\
J0717 & $7.09 \pm 0.39$ & $3.57 \pm 0.25$ & $2.30 \pm 0.21$ & $1.67 \pm 0.18$ & $1.24 \pm 0.15$
 \\
J0850 & $5.31 \pm 0.30$ & $1.87 \pm 0.20$ & $0.69 \pm 0.10$ & $0.32 \pm 0.07$ & $0.18 \pm 0.05$
 \\
J1149 & $4.96 \pm 0.35$ & $1.87 \pm 0.18$ & $0.89 \pm 0.12$ & $0.59 \pm 0.10$ & $0.41 \pm 0.09$
 \\
cluster average & $5.48 \pm 1.10$ & $2.22 \pm 0.77$ & $1.12 \pm 0.58$ & $0.73 \pm 0.44$ & $0.51 \pm 0.33$
 \\
\end{tabular}
\begin{tabnote}
\hangindent6pt\noindent
\hbox to6pt{\footnotemark[$*$]\hss}\unskip
These values are for a mock 5-year survey over a 1-$\mathrm{deg^2}$, with $\mathrm{t_{int}}=180~\mathrm{day}$ and $\mathrm{N_{d}}=2$.  The values for the cluster fields are extrapolated to the full 1-$\mathrm{deg^2}$ survey area.
\end{tabnote}
\end{table*}